\documentclass[10pt, conference]{IEEEtran}
\IEEEoverridecommandlockouts
\usepackage[left=0.6in,right=0.6in,bottom=0.98in, top=0.63in]{geometry}

\usepackage{cite}
\usepackage{multicol}
\usepackage{multirow}
\usepackage{hyperref}
\usepackage{amssymb}
\usepackage{float}
\usepackage{amsmath}

\allowdisplaybreaks
\usepackage{amsfonts}
\usepackage[utf8]{inputenc}
\usepackage[T1]{fontenc}
\usepackage{epstopdf}
\usepackage{algpseudocode}

\usepackage[english]{babel}
\usepackage[usenames, dvipsnames]{color}
\usepackage{pifont}
\usepackage{booktabs}
\usepackage{mathtools}
\usepackage{textcomp}
 \usepackage{graphicx}
\usepackage{subfigure}
\usepackage[ruled,vlined,commentsnumbered]{algorithm2e}
\usepackage{amsthm}
\usepackage{setspace}

\usepackage{eso-pic}

\SetKwInput{KwIn}{Inputs}

\newtheorem{proposition}{\textbf{Proposition}}

\theoremstyle{definition}

\theoremstyle{remark}

\theoremstyle{plain}

\newcommand{\RNum}[1]{\uppercase\expandafter{\romannumeral #1\relax}}

\def\BibTeX{{\rm B\kern-.05em{\sc i\kern-.025em b}\kern-.08em
    T\kern-.1667em\lower.7ex\hbox{E}\kern-.125em}}

\begin{document}
\title{Coexistence-Aware Antenna Tilt and Resource Allocation for Cellular-Connected UAV Corridors}
\author{\IEEEauthorblockN{Tarun Singh Rawat}
\IEEEauthorblockA{\textit{Department of Electrical Engineering} \\
\textit{Indian Institute of Technology Jodhpur}\\
Rajasthan, India\\
m25cps010@iitj.ac.in}
\and
\IEEEauthorblockN{Pradyumna Kumar Bishoyi}
\IEEEauthorblockA{\textit{Department of Electronics Engineering} \\
\textit{Indian Institute of Technology Jodhpur}\\
Rajasthan, India\\
pradyumna@iitj.ac.in}
}

\maketitle
\AddToShipoutPictureFG*{%
  \AtTextUpperLeft{%
    \put(0,\LenToUnit{0.3in}){\small This work has been accepted for publication in the IEEE ANTS 2026 Conference WS-08.}%
  }%
}

\begin{abstract}
Cellular-connected UAV corridors reuse terrestrial infrastructure to provide predictable aerial connectivity, but coexistence with ground users (GUEs) creates competing demands on antenna tilt and physical resource blocks (PRBs). Prior work optimizes tilt for signal quality alone and does not address joint tilt-PRB planning under corridor rate and GUE delay constraints, nor the allocation policy needed when combined resource claims exceed the PRB budget. We address both gaps by formulating a joint antenna-tilt and PRB-allocation problem that maximizes GUE sum-log-rate subject to a minimum UAV corridor rate and an $ M/G/1$-based GUE mean-delay bound. We decompose the problem into two subproblems: antenna tilts are adapted via projected gradient ascent, while PRB allocation follows a techno-economic approach. Under resource scarcity, we cast the PRB scarcity as a bankruptcy problem and apply the weighted Talmud rule to balance GUE and UAV claims. Simulation results demonstrate that relaxing the GUE delay bound consistently reduces UAV rate dissatisfaction under the proposed allocation.
\end{abstract}

\begin{IEEEkeywords}
UAV Corridor, Coexistence, PRB, FR1, NTN
\end{IEEEkeywords}

\section{Introduction} \label{sec:Introduction}
The growing demand for continuous and reliable uncrewed aerial vehicle (UAV) connectivity in logistics, inspection, surveillance, and urban mobility motivates research on extending cellular services to aerial users in upcoming 6G standard \cite{LAE_tut}. As UAV traffic increases, providing reliable connectivity throughout unrestricted three-dimensional airspace becomes difficult. In this context, cellular-connected UAV corridors have gained significant attention \cite{quantization_ICC,quantization_tcom,bayesian-optimization}. Such corridors confine UAV operations to predefined routes, thereby simplifying air traffic management and concentrating cellular coverage requirements within known aerial regions. Unlike a general cellular-connected UAV, whose trajectory is mission-dependent, or a UAV swarm, whose members coordinate their motion and tasks, UAVs sharing a corridor follow a spatially constrained route without requiring such coordination.

Reusing terrestrial cellular infrastructure for UAV corridors is cost-effective but creates coexistence challenges with existing cellular ground users (GUEs) \cite{grad_TVT,bishoyi_icc}. UAV corridors require a minimum downlink rate along predefined routes spanning multiple base stations (BSs). Two coupled design challenges then arise. First, down-tilted BS antennas provide weak aerial coverage, whereas uptilting antennas to improve corridor coverage strengthens interference over LoS links from neighboring BSs and can degrade GUE service \cite{harris_tcom}. Second, since UAVs and GUEs share the same limited pool of physical resource blocks (PRBs), congestion arises whenever their combined resource demand exceeds the available PRB budget. As antenna-tilt configuration and PRB allocation jointly determine the achievable rates for both user classes, cellular-connected UAV corridor systems require joint planning of the two to meet the service requirements of both. 


In \cite{quantization_ICC,quantization_tcom}, the authors employ quantization theory to optimize the vertical antenna-tilt of existing cellular base stations, with the objective of maximizing the average received signal strength (RSS) for both GUEs and UAVs. While \cite{bayesian-optimization} extends this approach via Bayesian Optimization to maximize the mean signal-to-interference-noise ratio (SINR) of both users. In \cite{dual-antenna}, the authors compare single- and dual-antenna designs for UAV corridor systems. Although these works improve UAV corridor connectivity, signal-quality optimization alone does not address joint antenna tilt and PRB planning under corridor rate and GUE delay requirements, nor does it provide an explicit allocation policy for when combined resource demands exceed the PRB budget, a gap that makes congestion management essential to network design.


In this work, we address two coupled network-design questions for cellular-connected UAV corridor systems. First, \textit{how should the BS antenna tilts be configured to guarantee the required rate along the corridor without significantly degrading GUE throughput?} Second, \textit{how should the limited PRBs be allocated to satisfy GUE delay requirements and corridor rate demands, particularly when their aggregate resource claims exceed the available budget?} To address these questions, we develop an economics-inspired resource-allocation framework that jointly optimizes antenna tilts and PRB allocations to maximize the sum of logarithmic GUE rates while imposing a minimum downlink rate at every sampled corridor location. We model each BS's aggregated GUE traffic as an $M/G/1$ queue and derive a closed-form service-rate threshold ensuring queue stability and the prescribed cell-level mean-delay bound. The service-rate threshold determines each BS’s minimum GUE PRB reservation. For each UAV corridor location, we derive the minimum PRB requirement from its downlink rate constraint. These quantities define the user's resource claims. When the available budget covers all claims, we satisfy them via a water-filling algorithm; otherwise, we cast the problem as a \textit{bankruptcy problem} \cite{thomson2012axiomatics}, and apply the weighted Talmud allocation (WTA) rule \cite{Talmud_economics}. The main contributions of our work are as follows,
\begin{itemize}
    \item We formulate a joint antenna-tilt and PRB-allocation problem to maximize the GUE throughput utility while ensuring UAV corridor minimum-rate requirements, GUE mean-delay constraints, and per-BS PRB budgets.
    \item We decompose the joint problem into antenna-tilt and PRB-allocation sub-problems, and solve them iteratively. The antenna-tilt subproblem is solved using projected gradient ascent with a logarithmic barrier to ensure UAV-rate feasibility. For fixed tilts, each BS uses water-filling when all minimum claims are feasible and WTA otherwise.
    \item Numerical results show that the fraction of UAVs failing to meet the minimum-rate requirement decreases under WTA as the GUE mean-delay constraint is relaxed. 
\end{itemize}

\section{System Model}\label{sec:SystemModel}
\begin{figure}
    \centering
    \includegraphics[width=0.46\textwidth]{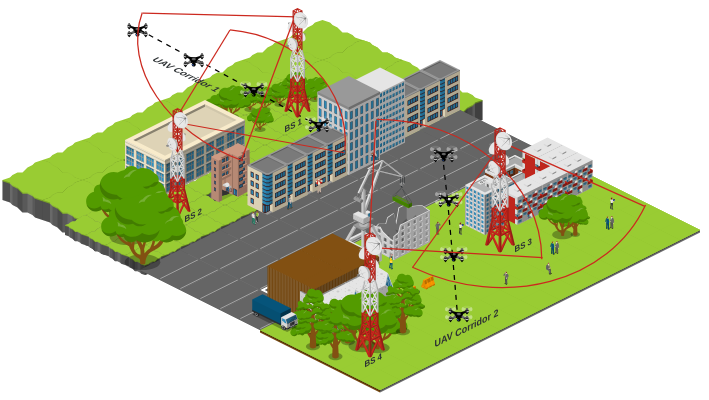}
    \caption{Illustration of Cellular-connected UAV corridor system.}
    \label{fig:blockdiagram}
\end{figure}

\subsection{Network Topology}
We consider a cellular network that consists of a set of $\mathcal{B}=\{1,2,\cdots B\}$ base stations (BSs) operating in frequency range 1 (FR1). The nominal inter-site distance (ISD) between neighboring BSs is denoted as $d_{ISD}$. We consider two types of users are being served by the BSs. First, the ground UEs (GUEs), indexed by $g\in\mathcal{E}_g$, are distributed uniformly around BS. Second, UAVs are inside the corridor at an altitude $h_a$ from the ground. The UAV locations in the corridor are indexed by $a\in\mathcal{A}$. The complete evaluation set is denoted as $\mathcal{E}=\mathcal{E}_g\cup\mathcal{A}$. 

Each BS $b \in \mathcal{B}$ provides service to both GUEs and UAVs using a finite number of physical resource blocks (PRBs). We consider a total of $N_b$ PRBs for BS $b$, and the per-PRB bandwidth is denoted as $B_{\mathrm{PRB}}$ (Hz). Further, the azimuth boresight direction $\psi_b$, and electrical tilt angle $\theta_b\in[\theta_{\min},\theta_{\max}]$. For a given tilt vector $\boldsymbol{\theta}$, let $\mathcal{U}_{b}^{g}(\boldsymbol{\theta})$, and $\mathcal{U}_{b}^{a}(\boldsymbol{\theta})$ denote the set of GUEs and UAVs associated with BS $b$, respectively. Further, $\mathcal U_b(\boldsymbol\theta)=\mathcal U_b^g(\boldsymbol\theta)\cup\mathcal U_b^a(\boldsymbol\theta)$, denotes all users associated with BS $b$.

\textit{Antenna gain}: The vertical and horizontal half-power beamwidths of the directional antennas are denoted by $\theta_{3dB}$ and $\psi_{3dB}$, respectively. The total antenna gain of BS $b$ in dB is given by $A_{b,q} \ = \ A_{max} \ + \ A_{b,q}^V  \ + \ A_{b,q}^H$,
where $A_{max}$ denotes the maximum antenna gain at the boresight, $A_{b,q}^V$ and $A_{b,q}^H$ represent the vertical and horizontal antenna gains in dB, respectively, given by $A_{b,q}^V \ = \ -\frac{12}{\theta_{3dB}^2}[\theta_{b,q} \ - \ \theta_b]^2$ and $A_{b,q}^H \ = \ -\frac{12}{\psi_{3dB}^2}[\psi_{b,q} \ - \ \psi_b]^2$. 
The vertical antenna gain $A_{b,q}^V$ depends on the vertical tilt $\theta_b$ of BS $b$ and on the elevation angle $\theta_{b,q}$ between BS $b$ and user location $q \in \mathcal{E}$, calculated as
\begin{equation}
    \theta_{b,q} \ = \ tan^{-1}\left(\frac{h_q \ - \ h_{b,\mathcal{B}}}{||q \ - \ p_b||}\right),
\end{equation}
where $h_q$, $h_{b.\mathcal{B}}$, $p_b$ represents height of user \textit{q} , height of BS $b$ and 2D coordinates of BS \textit{b} respectively. The horizontal antenna gain $A_{b,q}^H$ depends on the difference $[\psi_{b,q} \ - \ \psi_{b}] \ \in \ [-180^\circ,+180^\circ]$, where $\psi_b$ is azimuth orientation of BS \textit{b} and $\psi_{b,q}$ is the azimuth angle between BS $b$ and user location \textit{q}, given by
{\scriptsize
\begin{equation}
\psi_{b,q} =
\left\{
\begin{aligned}
&\tan^{-1}\left(\frac{q_y-p_{b,y}}{q_x-p_{b,x}}\right)
+ 180^\circ \times 2c,
&& if \ q_x-p_{b,x}>0, \\[6pt]
&\tan^{-1}\left(\frac{q_y-p_{b,y}}{q_x-p_{b,x}}\right)
+ 180^\circ \times (2c+1),
&& if \ q_x-p_{b,x}<0.
\end{aligned}
\right.
\label{eq:phi_bq}
\end{equation}
}
where subscripts $x$ and $y$ denote the horizontal and vertical
coordinates of a point, respectively, and the integer $c$ is chosen
in a way that $ -180^\circ \leq \psi_{b,q} \ - \ \psi_{b} \leq +180^\circ$.

\textit{Channel model}: We consider BSs are equipped with a directive antenna with a maximum gain of 14 dBi and a vertical (resp. horizontal) half-power beamwidth of $10^\circ$ (resp. $65^\circ$). All UEs are equipped with a single omnidirectional antenna. For notational compactness, $q\in\mathcal{E}$ denotes either a GUE $g$ or a UAV corridor location $a$. Let $G_{b,q}(\theta_b)$ denote the large-scale power gain from BS $b$ to user $q$, incorporating path loss, lognormal shadowing, and the antenna gain. Let $h_{b,e}$ denote the normalized small-scale fading coefficient. We assume Rayleigh fading for GUE links and pure LoS propagation for UAV links, modeled as $h_{b,g}\sim\mathcal{CN}(0,1),$ and $|h_{b,a}|^2=1$, respectively.


\subsection{Downlink SINR}
Assuming full frequency reuse, the downlink SINR of user $e\in\mathcal{U}_b(\boldsymbol{\theta})$ on a PRB is given by
\begin{equation}
\gamma_e(\boldsymbol{\theta})=\frac{P_b^{\mathrm{PRB}}G_{b,e}(\theta_b)|h_{b,e}|^2
}{
\displaystyle
\sum_{j\in\mathcal{B}\setminus\{b\}}P_j^{\mathrm{PRB}}G_{j,e}(\theta_j)|h_{j,e}|^2
+\sigma_{\mathrm{PRB}}^2
}.
\label{eq:downlink_sinr}
\end{equation}
Here, $P_b^{\mathrm{PRB}}$ denotes the transmit power per PRB at BS $b$, and $\sigma_{\mathrm{PRB}}^2$ is the noise power over bandwidth $B_{\mathrm{PRB}}$. Setting $e=g$ or $e=a$ yields the GUE SINR $\gamma_g(\boldsymbol{\theta})$ or the UAV SINR $\gamma_a(\boldsymbol{\theta})$, respectively.

\subsection{Downlink Communication Throughput}
We consider downlink communication throughput as a performance metric for both GUEs and UAVs. For GUEs, we consider the downlink log-sum rate, which is commonly used for maintaining fairness \cite{mine_1}. Given antenna tilt and PRB allocation, the log-rate is
\begin{equation}
U_g(\boldsymbol{\theta},\mathbf{n})
=
\sum_{g\in\mathcal{U}_b^g}
\log\!\left(
n_g B_{\mathrm{PRB}}
\log_2\!\left(1+\gamma_g(\boldsymbol{\theta})\right)
\right).
\label{eq:GUE_KPI}
\end{equation}

Similarly, the downlink communication rate along the UAV corridor points indexed by $a \in \mathcal{A}$ is
\begin{equation}
U_u(\boldsymbol{\theta},\mathbf{n})
=n_a B_{\mathrm{PRB}}\log_2\!\left(1+\gamma_a(\boldsymbol{\theta})\right).
\label{eq:UAV_KPI}
\end{equation}

\subsection{GUE delay modelling}
\label{subsec:aggregate_traffic_model}
We model the aggregate downlink GUE traffic at BS $b$ as an $M/G/1$ queue. Packets for GUE $g\in\mathcal{U}_{b}^{g}$ arrive according to an independent Poisson process with rate $\lambda_g$ packets/s and mean size $L_g$, yielding an offered load $\Lambda_g=\lambda_g\bar L_g$. By the superposition property of independent Poisson processes, the aggregate arrival rate and traffic load at BS $b$ are $\lambda_b^{tot}=\sum_{g\in\mathcal{U}_{b}^{g}}\lambda_g$ and $\Lambda_b=\sum_{g\in\mathcal{U}_{b}^{g}}\Lambda_g$, respectively. 
For fixed antenna tilts ($\theta$) and PRB allocations ($n_g$), the aggregate GUE downlink rate is
\begin{equation}\label{rate_mid}
	r_b=
	B_{\mathrm{PRB}}
	\sum_{g\in\mathcal{U}_{b}^{g}}	n_g\log_2\left(1+\gamma_g(\theta)\right),
\end{equation}
We approximate the aggregate queue as a constant-rate, first-come-first-served server with packet service time $S_b = L_b/r_b$. Under the stability condition $r_b>\Lambda_b$, the mean sojourn time including queueing and transmission can be expressed using the Pollaczek--Khinchine formula as
\begin{equation}
\begin{aligned}
	W_b
	&=
	\frac{\bar L_b}{r_b}
	+
	\frac{
	\Lambda_b\bar L_b
	\left(1+C_{L_b}^{2}\right)
	}{
	2r_b\left(r_b-\Lambda_b\right)
	}.
\end{aligned}
\label{eq:cell_sojourn_time}
\end{equation}
where $\bar L_b=(\sum_{g\in\mathcal{U}_{b}^{g}}\lambda_g\bar L_g
)/(\sum_{g\in\mathcal{U}_{b}^{g}}\lambda_g)$. $C_{L_b}^{2}$ denotes the squared coefficient of variation. 
Each BS must satisfy $W_b \leq D^{\mathrm{cell}}$, where $D^{\mathrm{cell}}>0$ denotes the prescribed cell-level mean-delay bound. The following proposition converts this requirement into a minimum service-rate condition.

\begin{proposition}
\label{prop:minimum_service_rate}
For a BS $b$ with $\Lambda_b>0$, the stability condition $r_b>\Lambda_b$ and mean-delay requirements $W_b\leq D^{\mathrm{cell}}$ are jointly satisfied if and only if
\begin{equation}
	r_b\geq r_b^{*},
\label{eq:cell_service_rate_requirement}
\end{equation}
where 
\begin{equation}
	r_b^{*}
	=
	\frac{
	D^{\mathrm{cell}}\Lambda_b+\bar L_b
	+
	\sqrt{
	\left[
	\begin{aligned}
		&\left(D^{\mathrm{cell}}\Lambda_b-\bar L_b\right)^2\\[-1mm]
		&\quad+
		2D^{\mathrm{cell}}\Lambda_b\bar L_b
		\left(1+C_{L_b}^{2}\right)
	\end{aligned}
	\right]
	}
	}{
	2D^{\mathrm{cell}}
	}.
\label{eq:minimum_service_rate}
\end{equation}
\end{proposition}

\begin{IEEEproof}
Under the queue-stability condition $r_b>\Lambda_b$, the term
$2r_b(r_b-\Lambda_b)$ is strictly positive. Substituting
\eqref{eq:cell_sojourn_time} into $W_b\leq D^{\mathrm{cell}}$ and multiplying $2r_b(r_b-\Lambda_b)$ both sides we get
\begin{align*}
	2\bar L_b(r_b-\Lambda_b)
	+
	\Lambda_b\bar L_b
	\left(1+C_{L_b}^{2}\right)
	\leq
	2D^{\mathrm{cell}}r_b(r_b-\Lambda_b).
\end{align*}
Rearranging the above inequality with respect to $r_b$ yields $D^{\mathrm{cell}}r_b^2
	-
	\left(
	D^{\mathrm{cell}}\Lambda_b+\bar L_b
	\right)r_b
	+
	\frac{
	\Lambda_b\bar L_b
	\left(1-C_{L_b}^{2}\right)
	}{2}\geq0$. 
Let $f_b(r_b)$ denote the left-hand side of the inequality. Since $D_{\text{cell}}>0$, 

$f_b$ is an upward-opening quadratic, and substituting $r_b=\Lambda_b$ gives
$f_b(\Lambda_b) = -\Lambda_b \bar{L}_b(1+C_{L_b}^2)/2 < 0$,  
so $\Lambda_b$ lies strictly between its two roots. Hence, within the stable region $r_b > \Lambda_b$, $f_b(r_b) \geq 0$ holds iff $r_b$ is no smaller than the larger root, which by the quadratic formula is $r_b^*$ in Eq. \eqref{eq:minimum_service_rate}. Since $r_b^{*}>\Lambda_b$, the condition
$r_b\geq r_b^{*}$ also guarantees queue stability.
\end{IEEEproof}

To translate the service-rate threshold in Proposition~\ref{prop:minimum_service_rate} into PRB requirements, we distribute the GUE reservations at BS $b$ in proportion to the offered traffic loads. We define each GUE's requirement as 
\begin{equation}
	c_g
	=
	\frac{\Lambda_g}{\Lambda_b}
	N_b^{\mathrm{GUE,min}},
\label{eq:gue_traffic_claim}
\end{equation}
A GUE with heavier traffic (i.e., a larger $\Lambda_g/\Lambda_b$) receives a proportionally larger reservation. Thus, the individual GUE requirements sum to the total reservation at BS $b$, i.e.,  $\sum_{g\in\mathcal{U}_{b}^{g}}c_g=	N_b^{\mathrm{GUE,min}}$. Under this allocation, we obtain the minimum reservation required to provide $r_b^*$ as
\begin{equation}
	N_b^{\mathrm{GUE,min}}
	=
	\left\lceil
	\frac{
	r_b^{*}
	}{
	B_{\mathrm{PRB}}\sum_{g\in\mathcal{U}_{b}^{g}}	\frac{\Lambda_g}{\Lambda_b}\log_2\left(1+\gamma_g\right)
	}
	\right\rceil.
\label{eq:minimum_gue_prbs}
\end{equation}



\subsection{Problem Formulation} \label{sec:ProblemFormulation}
Our goal is to determine optimal antenna tilt and PRB allocation that maximizes the aggregated sum-rate of GUEs while ensuring minimum downlink rate at each UAV corridor point and adhering to GUE mean-delay constraints. The optimization problem is formulated as
\begin{alignat}{2}
\textbf{P0}: \underset{\boldsymbol{\theta},\{n_g\},\{n_a\}}{\operatorname{maximize}}\quad
& U_g(\boldsymbol{\theta},\mathbf{n})
\nonumber\\
\operatorname{s.t}\quad
&U_u(\boldsymbol{\theta},\mathbf{n})\geq R_u^{\min},
\label{con:uav_rate}\\
& \sum_{g \in \mathcal{U}_b^g} n_g
+\sum_{a \in \mathcal{U}_b^a} n_a
\leq N_b, \forall\, b \in \mathcal{B},
\label{con:prb_budget}\\
& n_g\geq c_g,\quad g\in\mathcal{U}_{b}^{g}, \forall b\in\mathcal{B},\label{con:gue_prb}\\
& \theta_{\min}\leq\theta_b\leq\theta_{\max},\quad \forall b\in\mathcal{B},
\label{con:tilt_range}
\end{alignat}
Constraints \eqref{con:uav_rate}-\eqref{con:tilt_range}enforce the corridor rate, the per-BS PRB budget, the delay-driven GUE PRB floor, and the tilt range, respectively. Problem \textbf{P0} is nonconvex since both the utility functions ($U_g(\cdot)$ and $U_u(\cdot)$) are not jointly concave in the antenna tilts ($\boldsymbol\theta$) and PRB allocations ($n_g$). In Section III, we develop a solution framework for this coupled design and address resource allocation when the available PRBs cannot satisfy all minimum service requirements.


\section{Proposed Solution}\label{sec:ProposedMethod}
We adopt a two-step approach to solve the above optimization problem. We decouple the tilt-dependent SINR and PRB allocation by alternating between two subproblems until convergence. For fixed PRB allocations ($\{n_g\},\{n_a\}$), we optimize the BS tilts; for fixed tilts, we optimize the PRB allocation independently at each BS.


\subsection{Optimizing Antenna Tilt}
To decouple the antenna tilt ($\boldsymbol{\theta}$) from PRB allocation, we formulate a modified optimization problem that retains the tilt-dependent component of the objective in the problem \textbf{P0}. Further, the UAV rate constraint in Eq. \eqref{con:uav_rate} is incorporated through an equivalent spectral-efficiency feasibility threshold. 

Let $n_a^{max}$ be the maximum PRB that a BS can allocate to a UAV. We define the required spectral-efficiency threshold as $\tau_a = (R_u^{\min})/ (n_a^{max}B_{\mathrm{PRB}})$. Accordingly, UAV-rate feasibility during tilt optimization requires $\log_2\!\left(1+\gamma_a(\boldsymbol{\theta})\right) \geq \tau_a$. Clearly, $\tau_a$ ensures UAV link feasibility throughout the corridor. The modified objective function for optimizing antenna tilt in our considered scenario is
\begin{multline}
\Phi(\boldsymbol{\theta})=
\sum_{g\in\mathcal{E}_g}
\log\!\left(\log_2\!\left(1+\gamma_g(\boldsymbol{\theta})\right)\right)\\
+\kappa\sum_{a\in\mathcal{A}}
\log\!\left(\log_2\!\left(1+\gamma_a(\boldsymbol{\theta})\right)-\tau_a\right).
\end{multline}
The first term preserves the GUE utility in Problem \textbf{P0}, while the second is a logarithmic barrier enforcing UAV spectral-efficiency feasibility, with $\kappa>0$ controls the strength of the barrier. To this end, the antenna tilt optimization problem is formulated as,
\begin{alignat}{2}
\textbf{P1}:\quad  & \underset{\boldsymbol{\theta\in \Theta}}{\operatorname{max}}\quad
\Phi(\boldsymbol{\theta})
\nonumber
\end{alignat}
where $\Theta=[\theta_{\min},\theta_{\max}]^{|\mathcal{B}|}$ is the feasible set. We solve \textbf{P1} using projected gradient ascent \cite{boydconvex}. At iteration $t$, the antenna-tilt vector is updated as
\begin{equation}
    \boldsymbol{\theta}^{(t+1)}=\Pi_{\Theta}
\left[
			\boldsymbol{\theta}^{(t)}
			+\eta_t\nabla_{\boldsymbol{\theta}}
			\Phi\bigl(\boldsymbol{\theta}^{(t)}\bigr)\right]
\end{equation}
where $\eta_t > 0$ is the step-size, $\Pi_{\Theta}(\cdot)$ denotes
projection onto the feasible tilt set $\Theta$. The iteration continues till the difference between the $\boldsymbol{\theta}^{(t)}$ in two consecutive iterations is sufficiently small.

\subsection{Optimized PRB Allocation}
Now, in the second stage, we aim to find the optimal PRB allocation for GUEs and UAVs associated with each BS. For a fixed antenna tilt $\boldsymbol{\theta}^{*}$, PRBs are allocated independently at each BS. Therefore, the PRB allocation subproblem can be formulated as
\begin{alignat}{2}
\textbf{P2}:\quad  & \underset{\{n_g\},\{n_a\}}{\operatorname{max}}\quad
\sum_{g\in\mathcal{U}_b^{g}} \log(n_g) \quad \operatorname{s.t}\quad \eqref{con:uav_rate}, \eqref{con:prb_budget}, \eqref{con:gue_prb} \nonumber
\end{alignat}

Let $n_a^{min}$ denote the minimum PRB requirement for UAV. Then, from the Eq. \eqref{con:uav_rate}, the expression for $n_a^{min}$ can be expressed as $n_a^{\min}=\left\lceil\frac{R_u^{\min}}
{B_{\mathrm{PRB}}\log_2\!\left(1+\gamma_a^{*}\right)}\right\rceil$. 
Using Eq. \eqref{eq:minimum_gue_prbs} and $n_a^{min}$, the total PRB demand at BS $b$ can be expressed as
\begin{equation}
    F_b=N_b^{\mathrm{GUE,min}}+\sum_{a\in\mathcal{U}_b^{a}}n_a^{\min}
\end{equation}
Clearly, two regimes arise according to whether $F_b\leq N_b$, as explained in subsequent subsections.
\subsubsection{Feasible regime}
When $F_b \leq N_b$, the PRB budget of BS $b$ is sufficient to meet the minimum resource demand. We assign each UAV a minimum requirement $n_a^{\min}$ and allocate the remaining PRBs among GUEs using the traditional water-filling algorithm. The GUE allocation is $n_g^{*}=\max\left(c_g,\nu^{*}\right)$, where the optimal water level $\nu^{*}$  satisfies
\begin{equation}\label{nu_mid}
    \sum_{g\in\mathcal{U}_b^{g}}\max\left(c_g,\nu^{*}\right)=N_b-\sum_{a\in\mathcal{U}_b^{a}}n_a^{\min}
\end{equation}

\subsubsection{Resource-deficit regime}
When $F_b > N_b$, Problem \textbf{P2} is infeasible because all minimum claims cannot be simultaneously satisfied. We therefore model this using bankruptcy theory \cite{Talmud_economics}, which allocates a limited resource among users whose aggregate claims exceed its availability. We employ the weighted Talmud allocation (WTA) rule to account for both heterogeneous resource claims and service priorities \cite{Talmud_economics}. The WTA rule extends the classical Talmud allocation by incorporating user-specific priority weights alongside their resource claims. More specifically, the WTA is characterized by a triplet $( E_b, \boldsymbol{c}_b, \boldsymbol{\omega}_b), \forall b \in \mathcal{B}$,
where $E_b=N_b$ is the estate, $\boldsymbol{c}$ is the claim vector consisting of $c_g$, $\forall g \in \mathcal{U}_{b}^{g}$ and $n_a^{min}$, $\forall a \in \mathcal{U}_{b}^{a}$. Finally, $\boldsymbol{\omega}_b = (\omega_g^b, \omega_a^b)$ is the weight vector that represents the user's priority, where $\omega_g^b$ and $\omega_a^b$ are the weights assigned by BS $b$ for GUEs and UAVs, respectively.

\begin{algorithm}[ht]
\small
	\SetAlgoLined
	\SetKwInOut{Input}{Inputs}
	\SetKwInOut{Output}{Outputs}

	\Input{$\mathcal{U}_b^{g}$,$\mathcal{U}_b^{a}$,
	$\{c_g\}_{g\in\mathcal{U}_b^{g}}$,
	$\{n_a^{\min}\}_{a\in\mathcal{U}_b^{a}}$, and $N_b$}

	\Output{$\{n_e\}_{e\in\mathcal{U}_b}$ satisfying
	$\sum_{e\in\mathcal{U}_b}n_e\leq N_b$}

	Set $c_a\leftarrow n_a^{\min}$,
	for each UAV; Compute $F_b\leftarrow\sum_{e\in\mathcal{U}_b}c_e$\\

	\uIf{$F_b\leq N_b$}{\tcp{Water-filling allocation}
		Assign $x_a\leftarrow n_a^{\min}$,
		for each UAV\\
		Determine $\nu^{*}$ from Eq. \eqref{nu_mid} \\
		Assign $x_g\leftarrow\max(c_g,\nu^{*})$,
		for each GUE\\
	}
	\Else{
		\tcp{Weighted Talmud allocation}

		$E_b\leftarrow N_b$\\

		\uIf{$E_b\leq F_b/2$}{

			Find $\lambda$ by bisection using Eq. \eqref{eq:scarcity_budget} \;

			$x_e\leftarrow
			\min(c_e/2,\lambda\omega_e)$,
			$\forall e\in\mathcal{U}_b$\\
		}
		\Else{

			Find $\mu$ by bisection using Eq. \eqref{eq:loss_budget}\;

			$x_e\leftarrow
			c_e-\min(c_e/2,\mu/\omega_e)$,
			$\forall e\in\mathcal{U}_b$\\
		}
	}

	$n_e\leftarrow\lfloor x_e\rfloor$,
	$\forall e\in\mathcal{U}_b$.\\
	\caption{Per-Cell PRB Allocation}
	\label{alg:prb_allocation}
\end{algorithm}


The WTA rule is governed by the half-claim principle and operates in two regions depending on the available estate relative to one-half of the aggregated claim.
\begin{itemize}
    \item \textit{Scarcity region}: When $N_b \leq (F_b/2)$, the budget cannot cover more than half the aggregate claims. We allocate PRBs in proportion to the priority weights until individual allocations reach half their respective claims 
    \begin{equation}\label{eq:scarcity_budget}
      x_e = \min\bigg(\frac{c_e}{2},\lambda\omega_e\bigg)  
    \end{equation}
    where $\lambda$ is a tuning parameter that is chosen such that $\sum_{e \in \mathcal{U}_b} x_e = N_b$. 
    
    Before reaching the cap, a larger weight gives a user a larger allocation. The cap limits each user’s share relative to its own requirement.
    \item \textit{Relative-abundance region}: When $ F_b/2 < N_b < F_b$, the budget covers more than half the aggregate claims but remains insufficient to satisfy them fully. We therefore allocate the deficit $F_b-E_b$ as losses from the users’ claims:
        \begin{equation}\label{eq:loss_budget}
        x_e = c_e - \min\bigg(\frac{c_e}{2},\frac{\mu}{\omega_e}\bigg),
    \end{equation}
 The tuning parameter $\mu$ is chosen such that $\sum_{e\in\mathcal{U}_b}\min\left(\frac{c_e}{2},\frac{\mu}{\omega_e}\right)=F_b-E_b$. 
\end{itemize}
Here we can observe that the higher the value of $\omega_e$, i.e., the higher the priority of users, the smaller the resource loss in the relative-abundance region. The WTA relaxes (12) and (14) in problem \textbf{P0} while guaranteeing each user at least half its claim in the relative-abundance region. The detailed allocation rule and corresponding per-BS PRB allocation are summarized in Algorithm \ref{alg:prb_allocation}.

\textit{Complexity analysis}: Let $M_b=|\mathcal{U}_b|$ and $G_b=|\mathcal{U}_b^g|$. Water-filling algorithm requires
$\mathcal{O}(M_b+G_b\log G_b)$, where as WTA rule requires $\mathcal{O}(M_b I_b)$, where $I_b$ is the number of bisection iterations. 

\begin{figure}[t]
    \centering
    \vspace{1.5mm}
\includegraphics[width=0.36\textwidth]{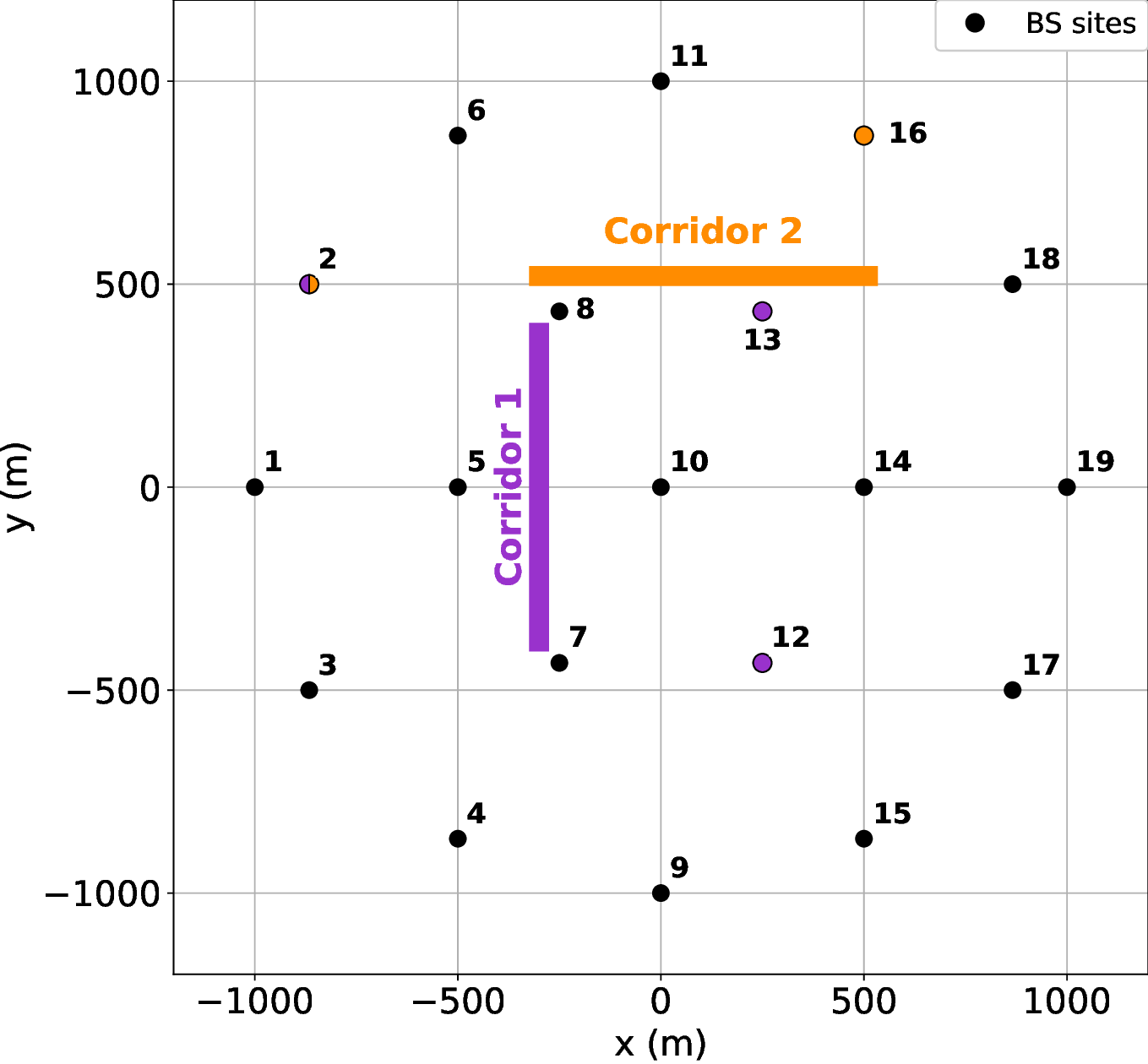} 
    \caption{A $19$-site deployment with $500$ m ISD and UAV corridors at 150 m and 120 m. Colors indicate the sites serving each corridor. Corridor 1 contains 5 UAVs and Corridor 2 contains 10 UAVs; sites are populated with 50 GUEs per BS.}
    \label{fig:System Model}
\end{figure}

\begin{figure}[ht!]
    \centering
    \vspace{1.5mm}
    \subfigure[GUE and UAV throughput under water-filling.\label{fig:cdf-watterfilling}]{%
  \includegraphics[width=0.8\linewidth]{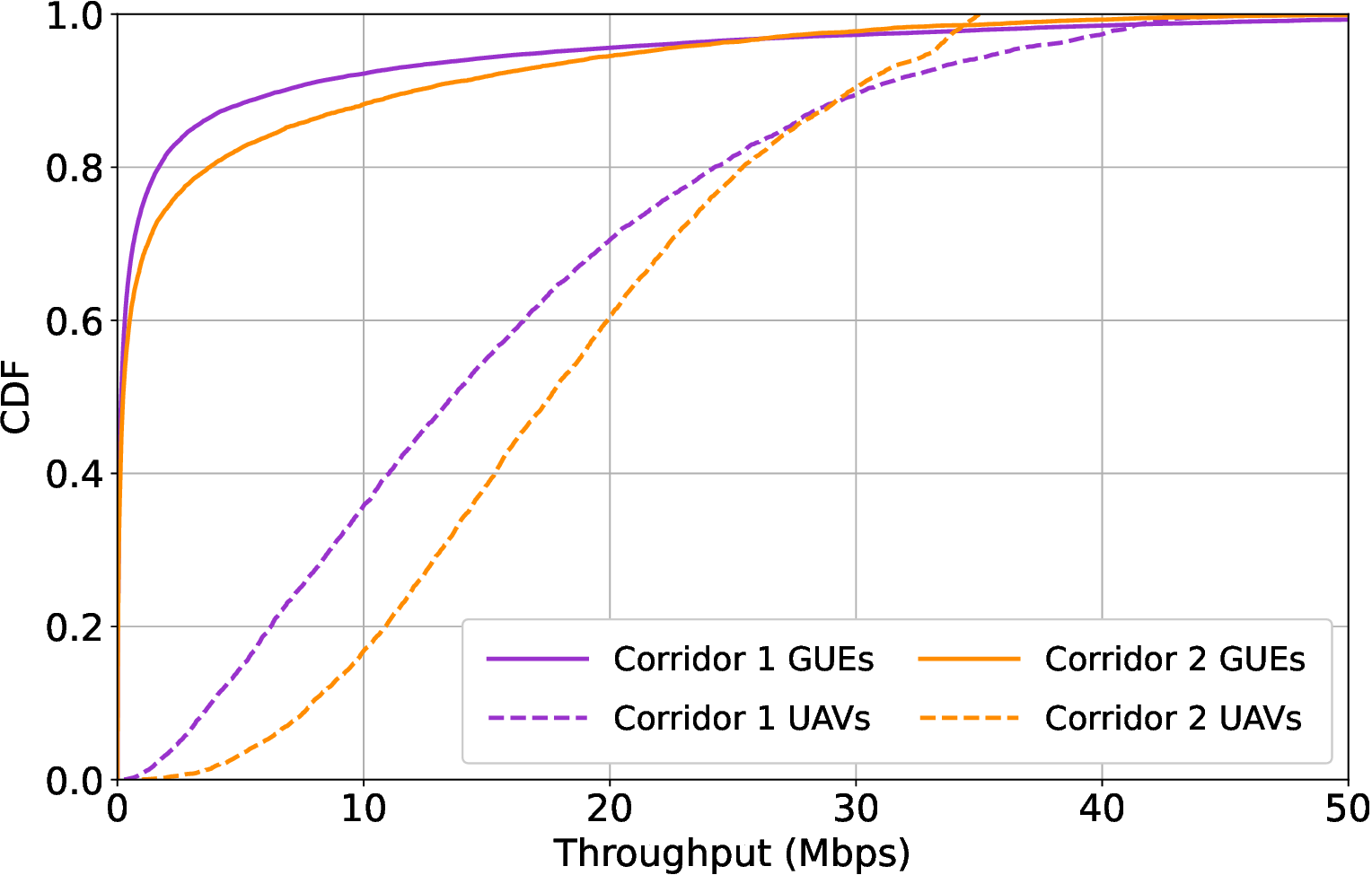}}
    \subfigure[GUE and UAV throughput under WTA rule.\label{fig:cdf-scarcity and abundane}]{%
        \includegraphics[width=0.8\linewidth]{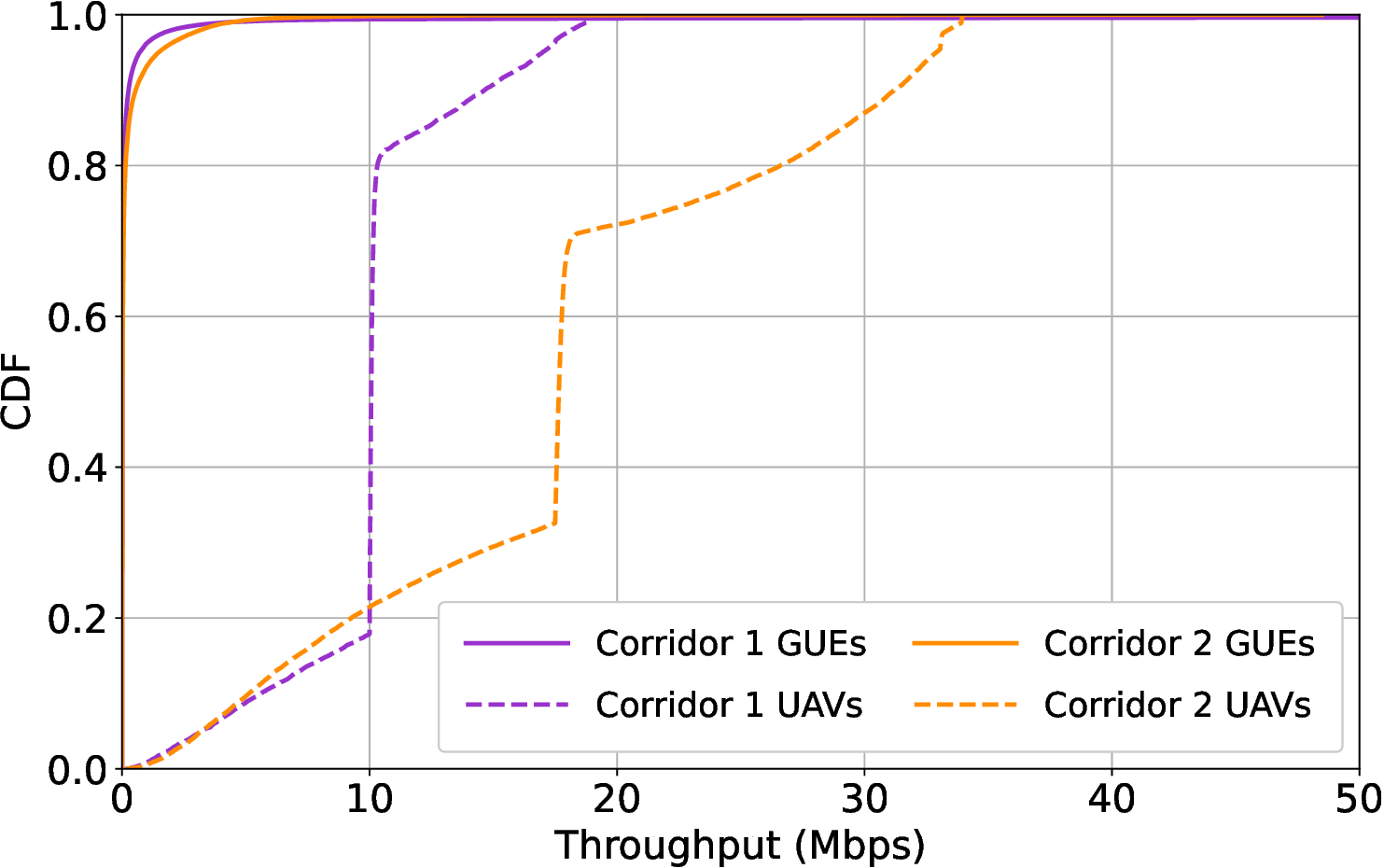}}
    \caption{Throughput CDFs for GUEs and UAVs in the two corridor scenarios: (a) water-filling; (b) WTA under PRB scarcity.}
      \label{fig:Throughput cdf} 
\end{figure}

\section{Performance Evaluation}\label{sec:Results}
\textit{Setup and Scenario}: We consider the cellular deployment in Figure \ref{fig:System Model}, with three sectors per BS oriented at $0^{\circ},120^{\circ}, \text{ and } 240^{\circ}$. We consider 5G NR FR1 carrier at $3.5$ GHz with a channel bandwidth of $100$ MHz and subcarrier spacing of $30$ KHz. Each BS has a PRB budget of $N_b = 275$ PRBs. The BS's transmitting power is set to $43$ dBm, uniformly distributed across its PRBs. Unless otherwise stated, we set $D^{cell} =50$ ms and $R_u^{min} = 10$ Mbps, $ \eta=0.02$, and barrier coefficient $\kappa=10$.

\begin{figure}[htbp]
    \centering
    \includegraphics[width=0.45\textwidth]{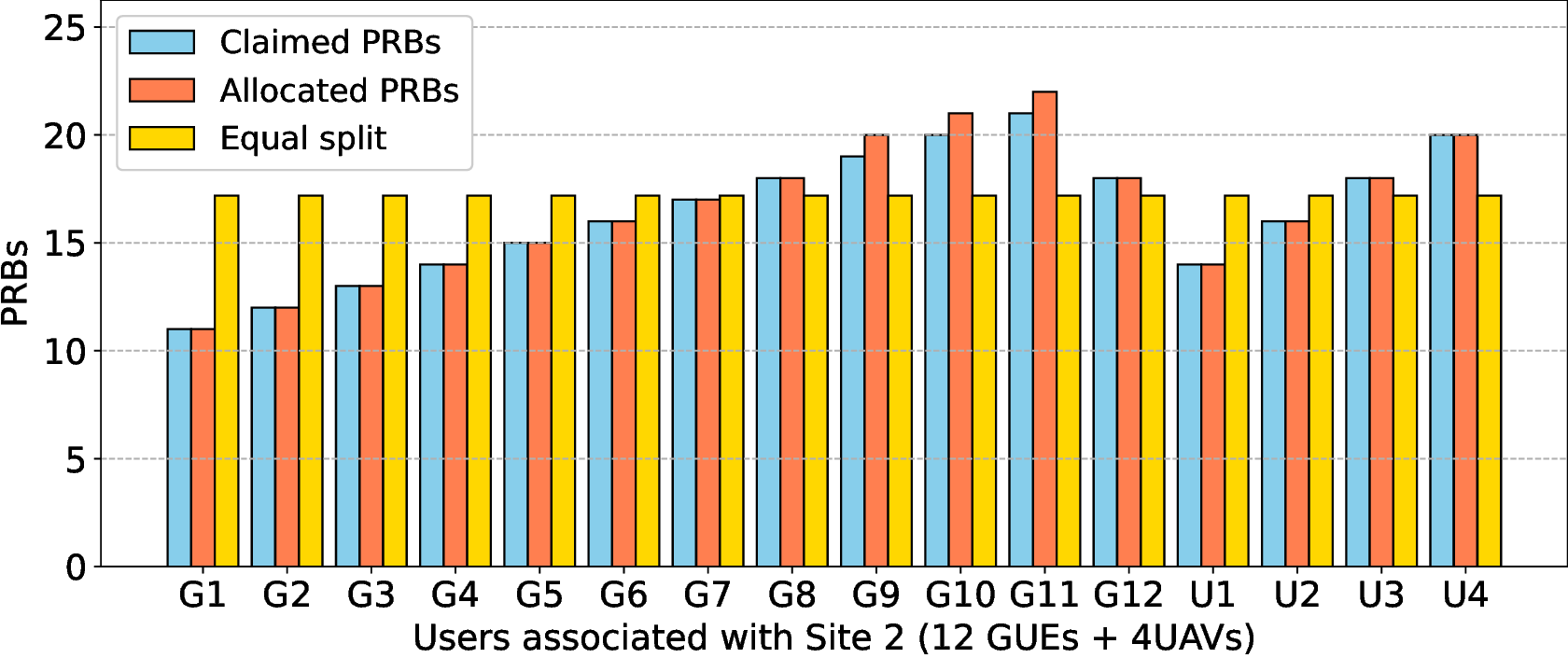} 
    \caption{PRB claims and allocations under the proposed method and equal sharing for 12 GUEs and four UAVs at Site 2.}
    \label{fig:PRB Allocation to users associated to Site 2}
\end{figure}
\textit{Simulation Results and Discussion}: Figure \ref{fig:Throughput cdf} presents the downlink throughput cumulative distribution functions (CDFs) of GUEs (solid line) and UAVs (dashed line) over $10^4$ Monte-carlo realizations under the two PRB-allocation regimes. Figure \ref{fig:cdf-watterfilling} shows the distribution under a feasible regime, i.e., the minimum resource demand is less than the BS PRB budget. We observe that the UAV throughput generally exceeds GUE throughput, with Corridor $2$ providing higher median UAV throughput than Corridor $1$. This is because the antenna-tilt favoring corridor UAVs reduces antenna gain towards GUEs. UAV reservations also limit the PRBs available to GUEs, whose service adequacy depends on the mean-delay requirement. 

Figure \ref{fig:cdf-scarcity and abundane} shows the distributions under WTA when aggregate claims exceed the PRB budget. UAV rates concentrate near $10$ and $18$ Mbps in Corridors $1$ and $2$, respectively, while GUE rates remain low. We observe that compared with Figure \ref{fig:cdf-watterfilling}, Figure \ref{fig:cdf-scarcity and abundane} exhibits a narrower UAV throughput range and stronger concentrations at specific rates. This is because the WTA distributes the insufficient budget according to claims and priorities. It therefore manages resource shortfalls without guaranteeing all service requirements are met. 


Figure \ref{fig:PRB Allocation to users associated to Site 2} compares individual PRB claims with the proposed allocation and equal sharing. We randomly select $12$ GUEs and four UAVs associated with Site 2, as shown in Figure \ref{fig:System Model}. Equal sharing assigns the same number of PRBs to each user, regardless of their requirements. It exceeds the claims of G1 and U1 but fails to satisfy G10 and U4. However, our proposed method meets all displayed claims before distributing surplus PRBs among GUEs. It achieves this by accounting for traffic-dependent GUE requirements and rate-dependent UAV requirements. 
\begin{figure}[ht]
    \centering
    \includegraphics[width=0.38\textwidth]{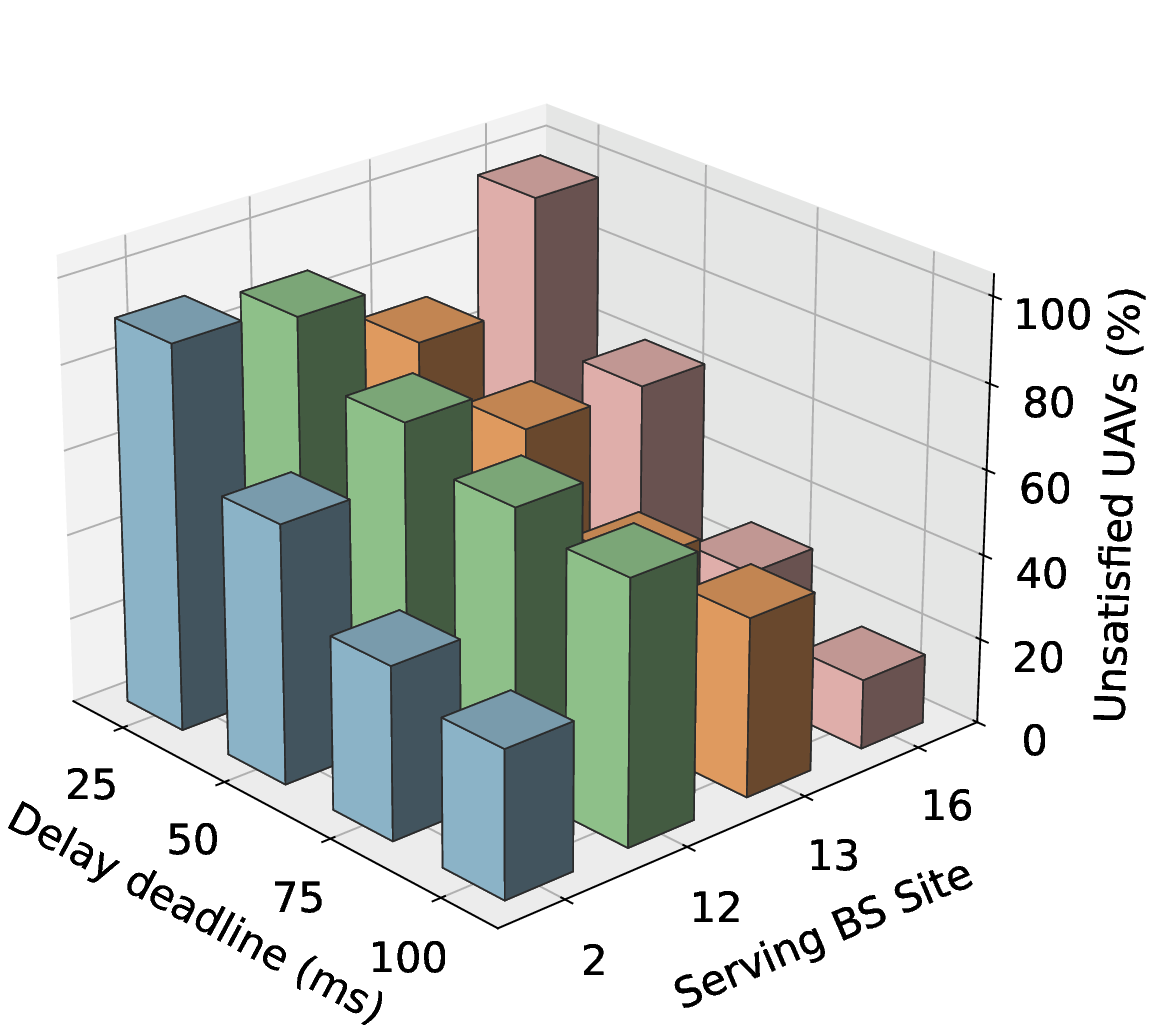} 
    \caption{Unsatisfied UAV percentage versus GUE mean-delay bound in the resource-deficit regime.}
    \label{fig:eps-example}
\end{figure}
Figure \ref{fig:eps-example} illustrates how GUE delay tolerance affects UAV rate satisfaction in the resource-deficit regime. A larger delay deadline value indicates applications that tolerate longer delivery times, rather than arbitrarily relaxing a fixed application requirement. Increasing it from 25 to 100 ms reduces UAV dissatisfaction at all four serving sites. This relaxation lowers the GUE service-rate threshold (Eq. \eqref{eq:cell_service_rate_requirement}) and corresponding PRB claims, allowing WTA to allocate more resources to UAVs. Thus, accommodating GUE application-specific delay requirements can improve UAV corridor service without increasing the PRB budget.
\begin{figure}[htbp]
    \centering
    \includegraphics[width=0.4\textwidth]{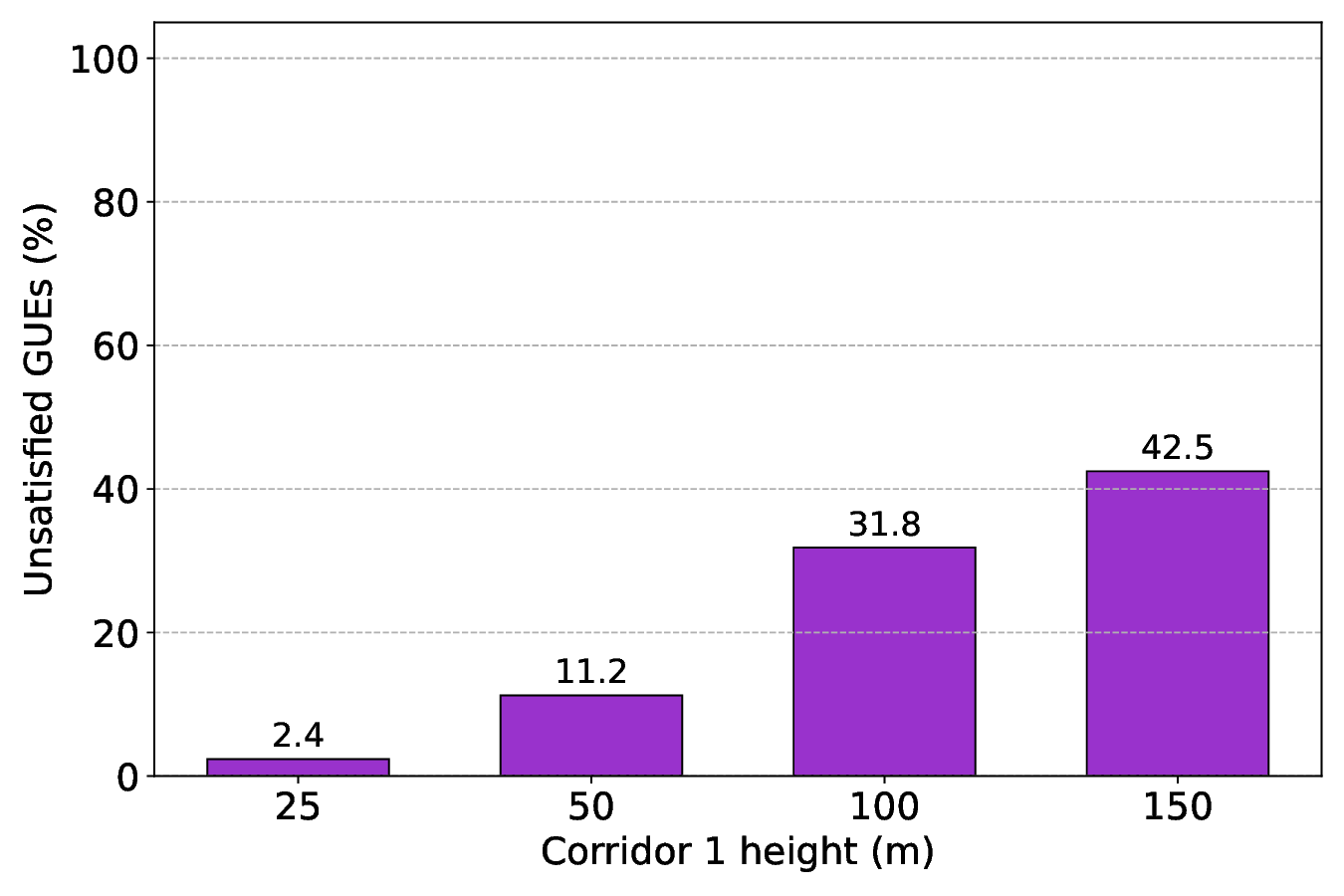} 
    \caption{Percentage of unsatisfied GUEs versus UAV corridor height.}
    \label{fig:unsatisfied GUE}
\end{figure}

Figure \ref{fig:unsatisfied GUE} shows the percentage of unsatisfied GUEs as the UAV corridor altitude is increased from 25 m to 150 m under the resource-deficit regime. Unsatisfied GUEs rise sharply from $2.4\%$ to $42.5\%$, as higher altitudes improve UAV line-of-sight to more BSs, inflating UAV PRB claims and biasing antenna tilts toward UAV coverage. This forces WTA rule to divert more PRBs away from GUEs, degrading their service. 




\section{Conclusions}\label{sec:Conclusions}
In this work, we proposed a joint antenna-tilt and PRB-allocation framework for cellular-connected UAV corridors, considering GUE mean-delay guarantees with UAV minimum-rate requirements under per-BS PRB budgets. Antenna tilt and PRB allocation are decoupled into two subproblems: tilt adaptation via projected gradient ascent and PRB allocation via water-filling when claims are feasible, or via a weighted Talmud rule under PRB scarcity. Numerical results show that WTA reduces UAV rate dissatisfaction as GUE delay tolerance is relaxed, demonstrating the efficacy of the proposed approach. 
\bibliographystyle{IEEEtran}
\bibliography{ref}

\end{document}